\documentclass[]{raa}            
\usepackage{graphicx,times}
\usepackage{psfig}
\usepackage{epsfig}
\usepackage{natbib}

\def\aap{A\&A}
\def\aaps{A\&AS}

\def\apj{ApJ}
\def\apjs{ApJS}
\def\apjl{ApJL}

\def\cas{Chinese Academy of Sciences}

\def\pasj{PASJ}

\def\raa{Research Astron. Astrophys.}

\def\solphys{Solar Phys.}

\def\los{line-of-sight}

\providecommand{\bm}[1]{\mbox{\boldmath$#1$\unboldmath}}

\def\etal{et al.}

\def\etc{etc .}
\def\los{line-of-sight}
\def\ie{i.e.}

\begin{document}

   \title{Statistical Study of Emerging Flux Regions and the Upper Atmosphere Response}

\volnopage{ {\bf 2012} Vol.\ {\bf XX} No. {\bf XX}, 000--000}
   \setcounter{page}{1}

 \author{Jie Zhao
      \inst{}
   \and Hui Li
      \inst{}
   }

   \institute{Purple Mountain Observatory, Nanjing 210008, China; \\
   {\it nj.lihui@pmo.ac.cn}\\ Key Laboratory of Dark Matter and Space Astronomy, CAS\\ }

   \date{Received~~2012 month day; accepted~~2012~~month day}

\abstract{We statistically study the property of emerging flux regions (EFRs) and the upper solar atmosphere response to the flux emergence by using data from the Helioseismic and Magnetic Imager (HMI) and the Atmospheric Imaging Assembly (AIA) on board the Solar Dynamics Observatory (SDO). Parameters including the total emerged flux, the flux growth rate, the maximum area, the duration of the emergence and the separation speed of the opposite polarities are adopted to delineate the property of the EFRs. The response of the upper atmosphere is addressed by the response of the atmosphere at different wavelengths (and thus at different temperatures). According to our results, the total emerged fluxes are in the range of (0.44 -- 11.2)$\times$10$^{19}$ Mx while the maximum area ranges from 17 to 182 arcsec$^2$. The durations of the emergence are between 1 and 12 hours, which are positively correlated to both the total emerged flux and the maximum area. The maximum distances between the opposite polarities are 7 -- 25 arcsec and are also correlated to the duration positively. The separation speeds are from 0.05 to 1.08 km s$^{-1}$, negatively correlated to the duration. The derived flux growth rates are (0.1 -- 1.3)$\times$10$^{19}$ Mx hr$^{-1}$, which are positively correlated to the total emerging flux. The upper atmosphere responds to the flux emergence in the 1600\AA\ chromospheric line first, and then tens and hundreds of seconds later, in coronal lines,
such as the 171\AA\ (T=10$^{5.8}$ K) and 211\AA\ (T=10$^{6.3}$ K) lines almost simultaneously, suggesting the successively heating of atmosphere from the chromosphere to the corona.
  \keywords{Sun: magnetic fields --- Sun: UV radiation --- Sun: corona
 }
  }

   \authorrunning{Jie Zhao \& Hui Li }
   \titlerunning{Statistical Study of Emerging Flux Regions and the Upper Atmosphere Response}

 \maketitle

\section{Introduction}           
\label{sect:intro}

The ubiquitous emerging flux regions (EFRs) on the Sun with variety of size, lifetime,
total magnetic flux and field strength have been widely discussed. Magnetic features with large scale
such as sunspots have fluxes of about 10$^{22}$ Mx (Maxwell) and generally exist in active regions (Thornton \& Parnell 2011). Magnetic features with small scale such as network fields and intranetwork (IN) fields have fluxes of 10$^{18}$ -- 10$^{19}$ Mx (Martin 1988; Wang et al. 1995) and 10$^{16}$ -- 10$^{18}$ Mx (Livingston \& Harvey 1975; Zirin 1985, 1987; Keller et al. 1994; Wang et al. 1995) respectively and generally exist in the quiet Sun. Using data from high-resolution Hinode (Kosugi \etal\ 2007) Solar Optical Telescope (SOT, Tsuneta \etal\ 2008), Thornton \& Parnell (2011) developed two different feature identification methods to determine the flux emergence rate of small-scale magnetic features in the quiet Sun. Combined with previous results, they found that the emergence frequency followed a power-law distribution with fluxes which ranged from 10$^{16}$ to 10$^{23}$ Mx . Simon \etal\ (2001) investigated the bipoles in the photosphere from emerging to splitting and then ending up in the magnetic network. They assumed a flux emergence rate of 7$\times$10$^{22}$ Mx d$^{-1}$ (consistent with that indicated by others for ephemeral regions (ERs)) that could keep the solar surface at a steady state. Hagenaar (2001) also examined a large number of ERs and obtained the total amount of flux emergence to be 5$\times$10$^{23}$ Mx d$^{-1}$ . They concluded that the magnetic field in the quiet Sun could be replaced in 14 hours with this emergence rate. When EFRs go through the atmosphere of the Sun, they produce various solar activities including ellerman bombs, blinkers, transient brightenings in small scale and solar flares, filament eruptions and coronal mass ejections (CMEs) in large scale as viewed by Low(1996) and heat the upper atmosphere (Li \etal\ 2007; Li \& Li 2010).

When a new EFR appears, it may interact with pre-existing surrounding region and produce a tiny two-ribbon flare (Sakajiri \etal\ 2004) or an EFR-surge which is the first signature of magnetic flux emergence in many EFRs (Kurokawa \& Kawai 1993). The evolution of two consequent dipoles in the coronal hole (CH) is first reported by Yang \etal\ (2009). The two dipoles interacted with each other and produced a jet and a plasma eruption. Their work is meaningful for the investigation of the CH evolution. Using the multi-wavelength observations combined with a nonlinear force-free extrapolation, Valori \etal\ (2012) provided a coherent picture of the emergence process of small-scale magnetic bipoles, which subsequently reconnected to form a large scale structure in the corona. Granular-scale flux emergence, which leaded to cancellation at the penumbral boundary was studied by Lim \etal\ (2011). They used data from the New Solar Telescope (NST, Goode \etal\ 2010) at Big Bear Solar Observatory (BBSO, Cao \etal\ 2010) with high spatial and temporal resolution. A bright point (BP) developed in their case due to the cancellation. They thought the scale of ER in their work was about 0.5 -- 1 arcsec, which was not detected in a magnetogram obtained with the Helioseismic and Magnetic Imager (HMI, Schou \etal\ 2011).

Hagenaar \etal\ (2008) investigated the evolution of magnetic network elements in the quiet-Sun photosphere with data from Michelson Doppler Imager (MDI, Scherrer \etal\ 1995) and found that the ER emergence rate is higher in flux-balanced regions. Wang \etal\ (2012) studied the solar IN magnetic elements. They found the flux emergence in these regions were mainly in the form of cluster emergence of mixed polarities and IN ERs. The samples in their work have an average separation of 3 -- 4 arcsec and lifetime of 10 -- 15 min, which are relatively small. Zhang \etal\ (2009) selected 6 events from Hinode Spectro-Polarimeter (SP, Lites \etal\ 2001) data to investigate the interaction between granulation and small-scale magnetic flux. Their result implies that the granule evolves quite differently according to the topology and emergence location of the EFR. Meanwhile, the granular flow also influences the development of EFR. With BBSO data, Zhang \etal\ (2006) compared the distribution of magnetic flux in a CH and a quiet region (QR). Their result demonstrates a balanced flux distribution in the QR and an imbalanced distribution in the CH, for IN fields and network fields . Yang \etal\ (2012) also statistically investigated the ERs in the quiet Sun and found two types of ERs: normal ERs and self-cancelled ERs. Their results also reveal that the ERs with higher magnetic flux tend to be self-cancellation easier. Statistical study about EFRs has also been done with SOT onboard the Hinode satellite by Otsuji \etal\ (2007, 2011). In the first paper they found the two polarities separated each other at a speed of 4.2 km s$^{-1}$ during the initial phase and then the separation speed decreased to about 1 km s$^{-1}$ ten minutes later. In the second paper, they demonstrated that the maximum spatial distance between two main polarities, the magnetic flux growth rate and the mean separation speed follow a power-law distribution with the total emerged flux. More works about magnetic fields can also be found in the review of Fang \etal\ (2011) and other references.

EFRs are probably the brightest features in the non-flaring solar corona (Schmieder \etal\ 2004). Responding to flux emergence, the coronal loops may appear bright in all temperatures. Yohkoh Soft X-ray Telescope (SXT, Tsuneta \etal\ 1991) has observed many transient brightenings (Shimizu \etal\ 1992, 1994) in multi-wavelength coordinated observations, which are located in EFRs. The close relation between the emerging flux and transient brightenings has been extensively studies (Mein \etal\ 2001; Kubo \etal\ 2003). Zhang \etal\ (2012) carried out a detailed multi-wavelength analysis of two coronal bright points (CBPs) and proposed that the gentle brightenings and the CBP flashes might be due to null-point reconnection and the separatrix reconnection, respectively.

Even though the SOT has observed many EFRs and some statistical work has been done to study EFR's properties (Otsuju \etal\ 2011), more work is still needed since the results are far from determined due to the wide span of their lifetimes, total fluxes, areas, \etc. Meanwhile, due to the intimate association of EFRs with various solar activities, statistical study of the properties of EFRs and the resultant response of the upper atmosphere is important to understand the physics of solar active regions and activities.

In this paper, we use data from the HMI and the Atmospheric Imaging Assembly (AIA, Lemen \etal\ 2011) on board the Solar Dynamics Observatory (SDO, Pesnell \etal\ 2011) to study the property of EFRs and corresponding response of the upper solar atmosphere. In Section \ref{sect:Obs}, we will introduce the observations and data reduction. We give one example to demonstrate our study as a case and the statistical results in Section \ref{sect:result}.
Our discussion and summary are presented in Section \ref{sect:sum}.

\begin{figure}  
\center{\resizebox{\textwidth}{!}{\includegraphics{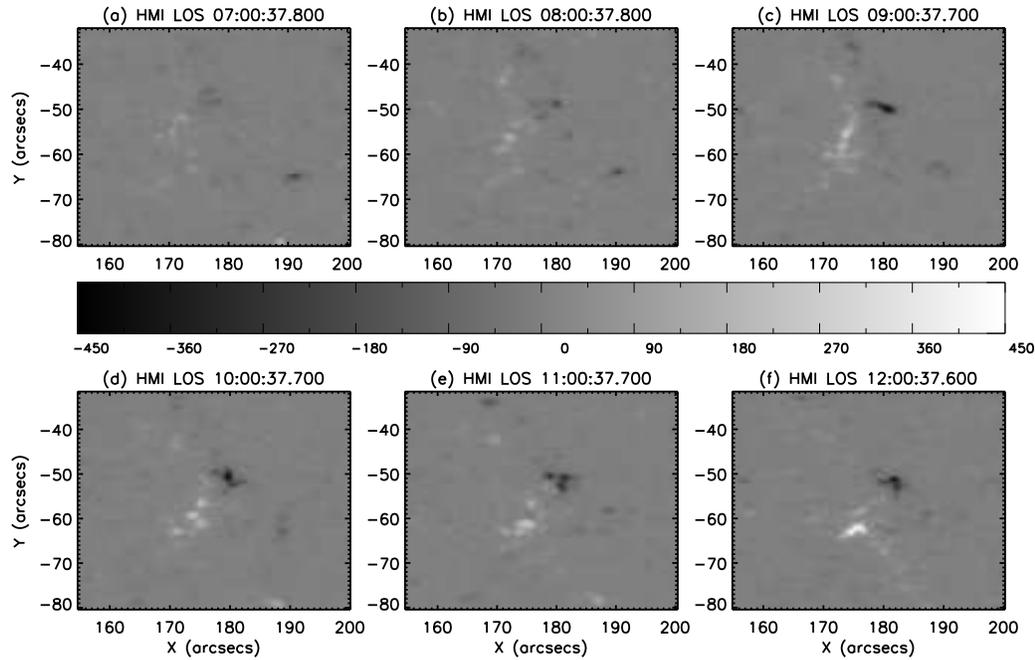}}}
\caption{LOS magnetograms obtained with SDO/HMI. It lasted from 2010 September 5 07:00 UT to 12:00 UT , which covers the time range of the studied EFR in the paper. The scale bar in the middle indicates the magnetic field strength in Gauss.}
\label{fig.hmi}
\end{figure}

\begin{figure}  
\center{\resizebox{\textwidth}{!}{\includegraphics{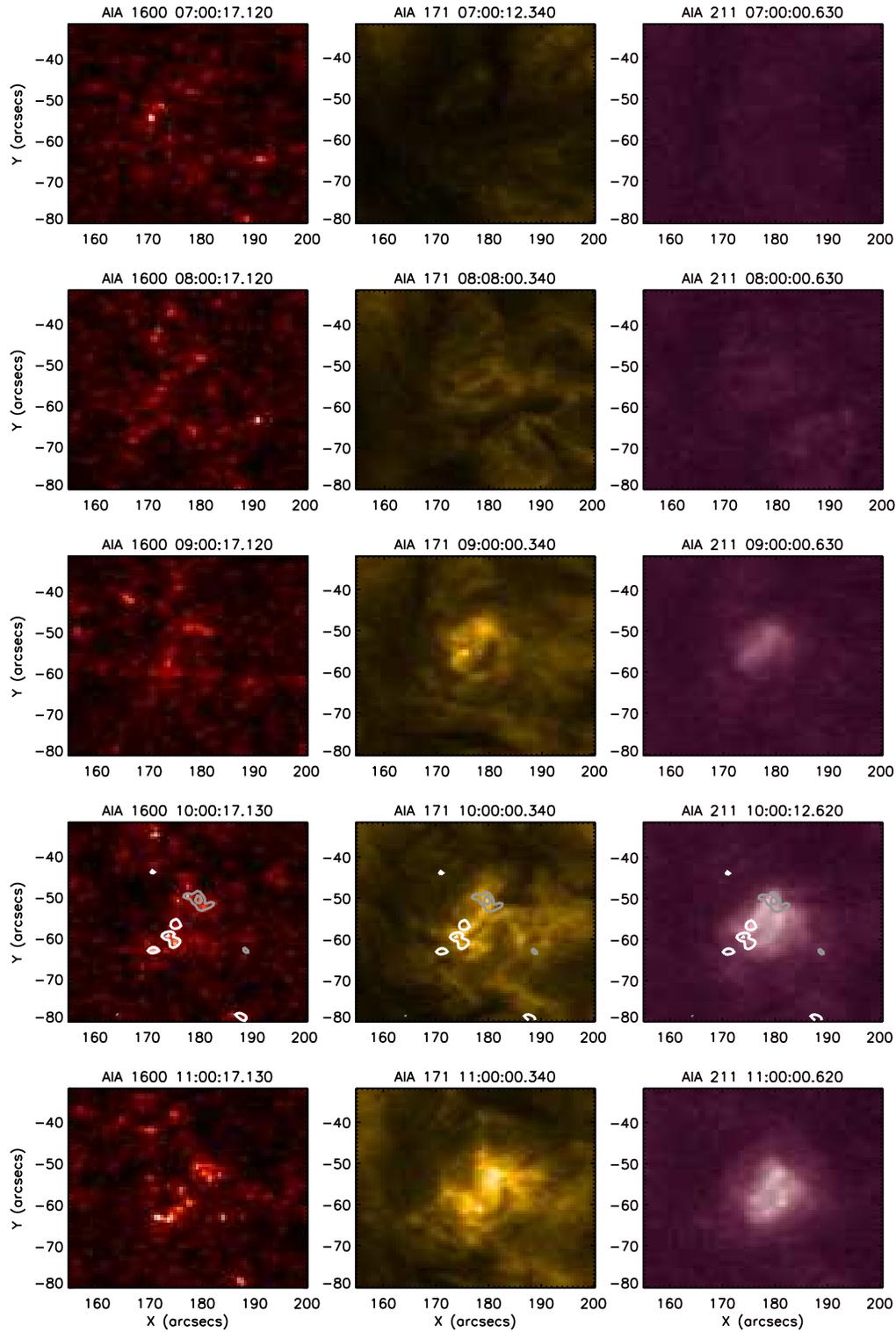}}}
\caption{Evolution of response in the upper atmosphere from AIA/1600\AA (left column), 171\AA (middle column), 211\AA (right column) lasts from 07:00 UT to 11:00 UT in the same region as in Figure \ref{fig.hmi}. The contours overlaid on the 10:00 UT panels are the LOS magnetic field with strength of 80, 250, -80, -250G. The white and gray lines correspond to positive and negative polarities, respectively. }
\label{fig.aia}
\end{figure}

\begin{figure}  
\center{\resizebox{\textwidth}{!}{\includegraphics{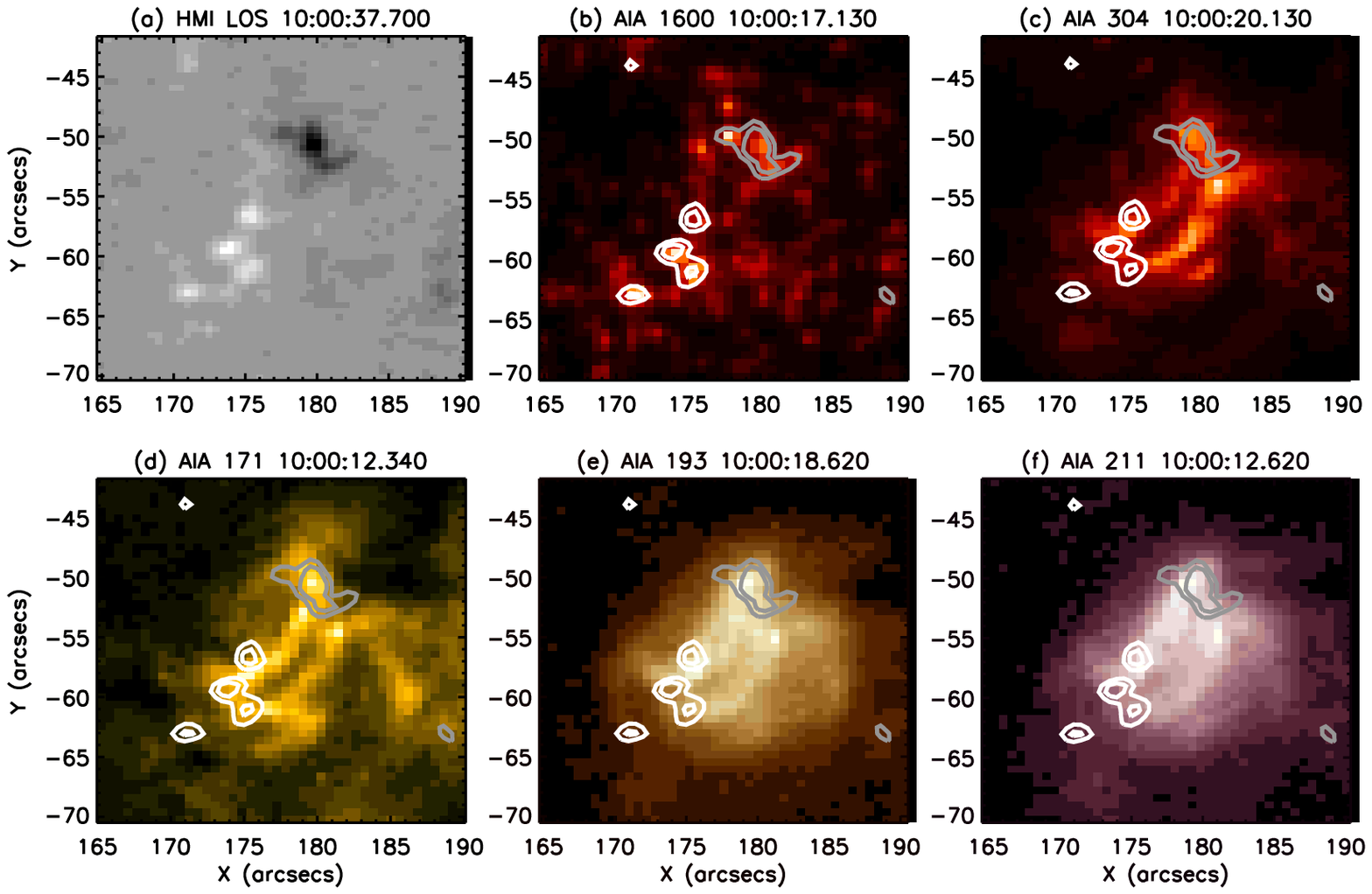}}}
\caption{LOS magnetogram and its corresponding AIA intensity maps in 1600\AA, 304\AA, 171\AA, 193\AA, and 211\AA, respectively. The region is a little smaller than the ones in Figure \ref{fig.hmi} and Figure \ref{fig.aia}. The contours overlaid on the intensity maps are the LOS magnetic field with strength of 80, 160, -80, -160G. The white and gray lines correspond to positive and negative polarities, respectively.}
\label{fig.over}
\end{figure}

\section{Observations And Data reduction }
\label{sect:Obs}

We use the iSolSearch tool of the Heliophysics Events Knowledgebase (HEK) system (http://www.lmsal.com/isolsearch) to select EFRs for our study, which allows us to easily find EFRs and download data for the required field-of-view (FOV) to save time and computer disk space. The studied EFRs are selected according to the following three criteria: (1) they appear close to the disk center to minimize the project effect of the coronal observation, namely, the response in extreme ultra-violet (EUV) wavebands; (2) they are relatively simple and without interaction with other EFRs to guarantee the
accuracy of the computed separation and subsequently the separation speed (see definition below for these parameters); (3) they keep on emerging for at least one hour. Based on these requirements, we selected 50 EFRs to conduct this study, which appeared on 2010 September 2 -- 6, 2011 August 27 -- 29, 2011 September 10 and 2011 October 21, respectively.

The HMI instrument on SDO observes the full solar disk in the Fe I absorption line at 6173 \AA\ with a resolution of 1 arc-second. It provides four main types of data: dopplergrams (maps of solar surface velocity), continuum filtergrams (broad-wavelength photographs of the solar photosphere), and both \los\ (LOS) and vector magnetograms (maps of the photospheric magnetic field). The AIA onboard SDO obtains full-sun images in multiple EUV and ultra-violet (UV) passbands with a resolution of 1.2 arcsec. The two instruments provide the first full-disk continuous observations of solar magnetic fields and solar atmosphere, respectively. HMI obtains the LOS magnetic field with a 45 s cadence and AIA records coronal images with a 12 s cadence.

In our study, we adopt 12 min HMI LOS field data. Considering the magnetic field evolves slow with respect to the duration of EFR, data with such a cadence is acceptable to describe the property of EFR. The downloaded HMI magnetic field data have been calibrated. We download AIA images in partial form based on the region of EFR in, such as 1600 \AA, 304 \AA, 171 \AA, 193 \AA, and 211 \AA\ lines, which are formed at the upper chromospheric and the coronal temperatures. The downloaded AIA images are prepared in a standard manner including bad-pixel removal, despiking and flat-fielding. Both HMI and AIA data are corrected for the differential rotation. The HMI and AIA images are coaligned by coaligning the HMI intensity image and AIA white-light (WL) image.

To study the property of EFRs, we derive the following parameters from HMI LOS magnetic field data: (1) emerging duration (time span from the start to the maximum flux), (2) total emerged flux (maximum flux subtracted by the flux before emergence), (3) flux growth rate (total emerged flux divided by the emerging duration), (4) maximum area (the maximum value for the area of EFR), (5) average field strength (arithmetic mean value of the field strength, which is calculated through total emerged flux divided by the area), (6) separation (distance between the two opposite polarities), and (7) separation
speed (fitting result from the time profile of the separation).

\section{Results}
\label{sect:result}

\begin{figure}  
\center{\resizebox{\textwidth}{!}{\includegraphics{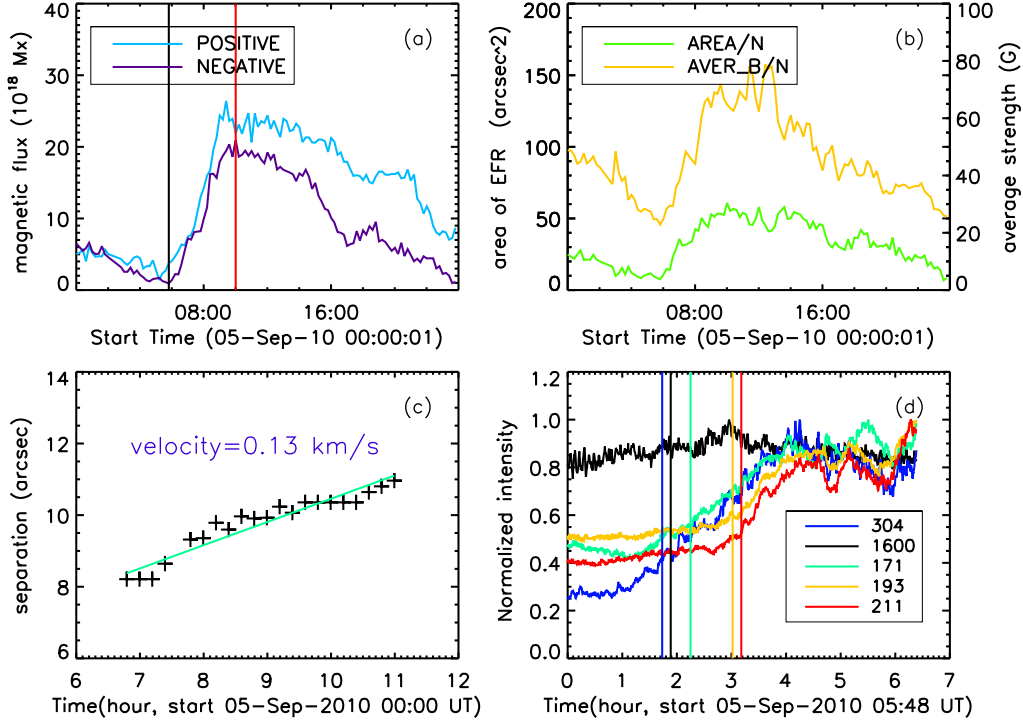}}}
\caption{Panel (a) shows the evolution of magnetic flux in the region which shown in Figure \ref{fig.over} (a). The vertical lines indicate the start and end time of the emergence, respectively. Panel (b) gives the area and average magnetic field strength variations of the EFR in the same region as panel (a). Evolution of distance between the opposite polarities of EFR is shown in panel (c). The green line indicates the fitting result. We exhibit light curves of 5 AIA channels in panel (d) which all start from 2010 September 5 05:48 UT. The vertical lines indicate the start times of response in the same colors. }
\label{fig.flux}
\end{figure}

\subsection{Example}
We present the event happened on 2010 September 5 as a case study. The evolution of this EFR is shown in Figure \ref{fig.hmi}. This EFR appeared before 07:00 UT and we can see the two polarities(even they are weak) located at the upper-left corner from center in panel (a). In this rectangular region, a single negative pole exists at the lower-right corner and we can still see it in panel (b). However, this rectangular area is relatively clear from 09:00 UT since the EFR has developed and become the main feature in this region. The distance between opposite polarities increased obviously after 07:00 UT and changed slightly after 10:00 UT. The magnetic dipole seems to move slightly from upper-left to lower-right, which may suggest that the convective flow is coupling with the magnetic field during flux emergence.

Images from three of the five selected AIA channels are presented in Figure \ref{fig.aia}. The AIA/1600 \AA, 171 \AA, 211 \AA\ intensity maps are shown in the left column, middle column and right column, respectively. All of them are recorded from 07:00 UT to 11:00 UT. The region has the same FOV as the rectangular area in Figure \ref{fig.hmi} and three intensity maps at 10:00 UT are overlaid with contours of the magnetic field at the same time. The contours manifest that the magnetic dipole corresponds to the chromospheric brightening features and footpoints of the coronal loops. In 171 \AA\ and 211 \AA\ intensity maps, coronal loops become larger and brighter with time although we can just see two obvious brightening features in 1600 \AA\ intensity maps.

We show the LOS magnetogram together with five intensity maps of AIA in Figure \ref{fig.over} with the same area in which we calculate the parameters mentioned in section \ref{sect:Obs}. The contours overlaid on the intensity maps correspond to the magnetic dipole.

The area that we select to calculate the parameters mentioned above is of proper size in order not to include other magnetic structures as shown in Figure \ref{fig.hmi} (a). We use $\pm$20 G as the background magnetic field when computing the magnetic flux of the EFR, which is shown in Figure \ref{fig.flux} (a). It clearly manifests an emerging event in this region, which started before 06:00 UT and reached the maximum at about 10:00 UT as indicated by the vertical lines. The start time of emerging is defined to be the time when the magnetic flux increases continuously while the end time of emerging is chosen to be the time when the magnetic flux reaches the maximum. The span of this two times is defined as the duration of EFR. The area of EFR is obtained by summing all pixels with magnetic field strength larger than 20 G or less than -20 G . With this threshold, we determine the average magnetic field strength in the EFR (Figure \ref{fig.flux} (b)). It only gives the area and average magnetic field of negative pole of EFR. We should mention that even we are trying to choose a region that only contain the EFR we are interested in, the region may have other magnetic features more or less and subsequently affect the result, such as magnetic flux of the EFR. Hence, we will pick the polarity with smaller maximum magnetic flux to represent the whole EFR. For this example, maximum magnetic flux of negative polarity is smaller, so it is more suitable to represent the EFR.

Using the IDL program( 'label-region.pro' ), we can label connected domain in this rectangular area, then the distance between the two polarities is calculated and shown in Figure \ref{fig.flux} (c).We show the distance variation after 06:00 UT, but keep in mind that the EFR appeared before 06:00 UT although it is faint in the magnetogram. Since the EFR is very small and faint in magnetogram at the beginning, our method may capture other features, which is more obvious at the that time. Only when the EFR becomes the main features in the region, the derived distance is what we need. So we ignore the initial phase when computing the separation speed if there are other features. The separation speed is easy to obtain by fitting the evolution curve of distance between the two polarities. For this EFR, a velocity of 0.13 km s$^{-1}$ is obtained.

From SDO/AIA data, we got light-curves of the five layers from chromosphere to corona, shown in Figure \ref{fig.flux}(d). The curves all start from 2010 September 5 05:48 UT when the emergence started. All of them increase continuously after the EFR appears and the light curves with higher temperature seem to start increasing later. The background values of the five channels are the average intensities from 05:48 UT to 06:30 UT. When the value reaches 20\% of the maximum increment, we define this time as the response time of atmosphere.

\begin{figure}  
\center{\resizebox{\textwidth}{!}{\includegraphics{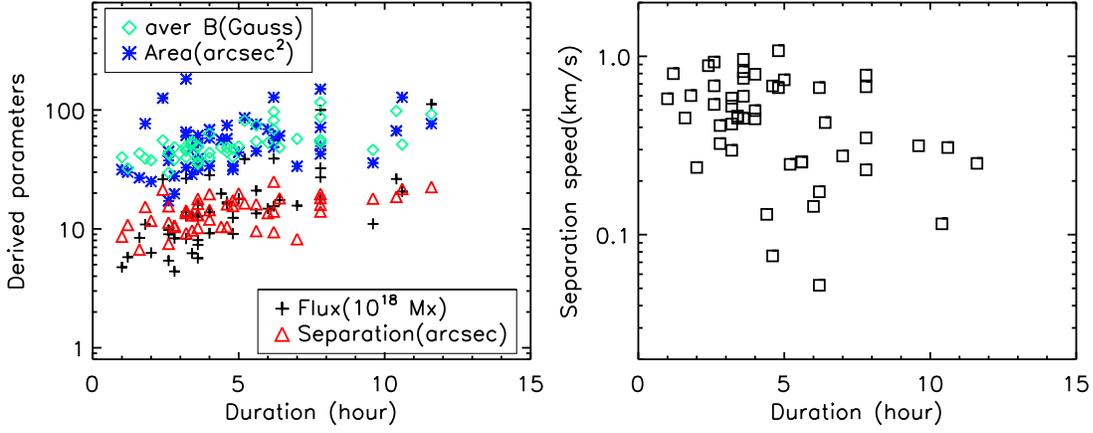}}}
\caption{Distributions of the derived emerged flux, maximum area, separation and average field strength (left) and the separation speed (right) with the emerging duration for the studied EFRs. }
\label{fig.mag}
\end{figure}

\begin{figure}  
\center{\resizebox{\textwidth}{!}{\includegraphics{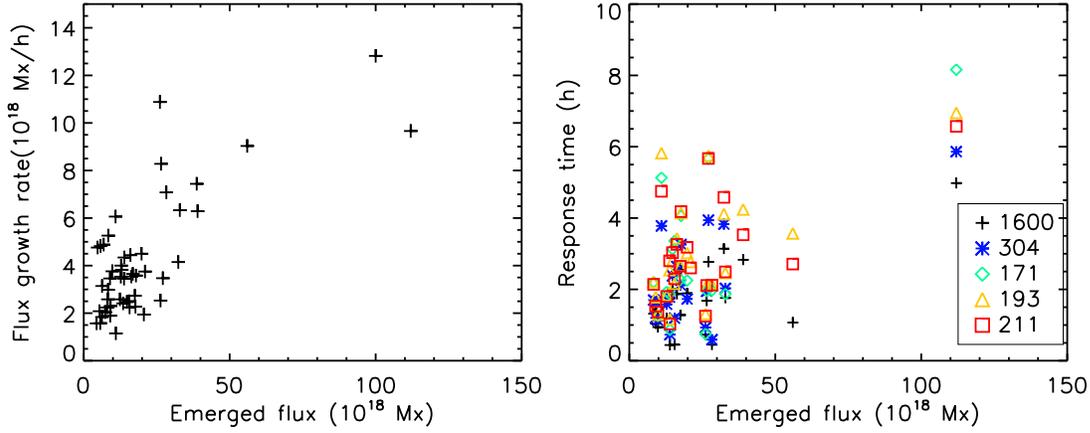}}}
\caption{Distributions of flux growth rate (left) and response time of different AIA wavelengths with respect to start time of flux emergence (right).}
\label{fig.resp}
\end{figure}

\begin{table}
\begin{center}
\caption[]{The basic information and derived parameters for the studied EFRs. }
\label{Tab:EFRs}


 \begin{tabular}{ccccccccc}
  \hline\noalign{\smallskip}
 EFR  & Polarity & Flux &  Duration & Emerging rate & Area & Aver B  & Separation & Sep speed\\
 number &     & (10$^{19}$ Mx) & (hour) &(10$^{18}$ Mx hr$^{-1}$) & (arcsec$^2$) & (G)& (arcsec)&(km s$^{-1}$)\\
 \hline\noalign{\smallskip}
 20110827C & + &0.44&2.8&1.57&19.7&38.2&10.7&0.41\\
 20100902E & -&0.48&1.0&4.76&31.5&40.1&8.6&0.58\\
 20100905C & -&0.54&2.6&2.08&17.2&29.7&7.5&0.54\\
 20100902F & -&0.57&3.6&1.58&35.5&36.5&17.9&0.96\\
 20110827A & +&0.58&1.2&4.82&30.0&32.3&10.8&0.8\\
 20100902D & -&0.62&3.4&1.84&28.7&54.4&9.7&0.46\\
 20110827E & -&0.63&2.0&3.14&25.0&37.8&11.7&0.24\\
 20110827D & -&0.68&1.4&4.88&26.4&30.6&-&-\\
 20100902B & +&0.74&3.6&2.04&31.4&35.2&10.2&0.75\\
 20100902G & -&0.81&3.6&2.24&38.8&49.9&13.1&0.6\\
 20110827G & -&0.83&3.2&2.58&32.8&39.9&9.2&0.42\\
 20100903F & -&0.83&2.8&2.98&27.8&48.7&10.4&0.32\\
 20100902C & -&0.84&1.6&5.26&26.9&43.2&6.7&0.45\\
 20111021E & -&0.90&2.6&3.46&30.7&42.6&11.3&0.93\\
 20110827B & +&0.91&4.8&1.89&33.7&40.4&15.6&0.67\\
 20100906C & -&0.92&4.0&2.3&33.9&43.7&19.6&0.79\\
 20100903D & +&0.98&2.6&3.76&37.6&43.2&15.5&0.68\\
 \hline\noalign{\smallskip}
 20110829A & +&1.1&1.8&6.06&76.7&39.1&15.3&0.6\\
 20100905A & -&1.1&9.6&1.15&35.9&46.3&18&0.32\\
 20100905F & -&1.24&4.8&2.58&31.5&47.2&17.4&1.08\\
 20100903E & -&1.27&3.6&3.53&52.9&48.9&13.1&0.45\\
 20110910E & +&1.28&3.2&4.0&42.7&45.4&13.6&0.58\\
 20100903B & -&1.3&3.4&3.82&48.4&51.7&13&0.45\\
 20110827I & +&1.35&5.6&2.41&45.1&54.5&9.6&0.26\\
 20100902H & +&1.38&4.0&3.45&58&39.7&12&0.45\\
 20110910D & -&1.39&3.2&4.34&76.1&46.3&13.8&0.3\\
 20110828B & -&1.48&6.0&2.47&69&53.2&13.6&0.14\\
 20100903C & -&1.55&6.2&2.5&48.8&68.3&9.4&0.05\\
 20110827J & +&1.57&7.0&2.24&33.6&57.3&8.2&0.28\\
 20110829B & +&1.6&3.6&4.44&60.9&40&16.3&0.82\\
 20100906D & -&1.63&4.6&3.54&57.3&44.3&16.1&0.68\\
 20110827F & +&1.68&4.6&3.65&74.2&48.1&10.4&0.08\\
 20100903A & -&1.75&6.4&2.73&60.6&48.4&18.1&0.43\\
 20110828A & +&1.77&7.8&2.27&71.4&53.1&19.7&0.79\\
 20100905B & +&1.79&5.0&3.58&44.1&49.5&19.9&0.74\\
 20100905D & -&1.98&4.4&4.5&56.1&48.4&10.4&0.13\\
 20110829C & -&2.06&10.6&1.94&66.8&51.3&21.7&0.31\\
 20111021B & +&2.1&5.6&3.75&44.8&74.4&16.1&0.26\\
 20100906A & +&2.61&2.4&10.88&125.7&55.6&21.4&0.89\\
 20110910C & +&2.63&10.4&2.53&149.9&98.2&18.6&0.11\\
 20100902A & +&2.65&3.2&8.28&65.3&49.7&14.3&0.52\\
 20100905E & +&2.71&7.8&3.47&49.8&56.2&14&0.23\\
 20100906B & -&2.83&4&7.08&68.4&63.2&14.4&0.5\\
 20111021C & +&3.24&7.8&4.15&80.3&86.9&16.0&0.68\\
 20111021D & -&3.29&5.2&6.33&87.1&56.9&-&-\\
 20110827H & -&3.87&5.2&7.44&85.9&81.2&16.4&0.25\\
 20110828C & +&3.9&6.2&6.29&127.7&81.8&25&0.68\\
 20110910B & -&5.6&6.2&9.03&61.2&96.8&14&0.17\\
 \hline\noalign{\smallskip}
 20111021A & +&10&7.8&12.8&111.4&116.1&18.2&0.35\\
 20110910A & -&11.2&11.6&9.66&182.2&92&22.5&0.25\\

 \noalign{\smallskip}\hline
\end{tabular}
\end{center}
\end{table}

\begin{table}
\begin{center}
\caption[]{ Time delay of the selected five AIA channels with respect to the start time of flux emergence. }
\label{Tab:response}


 \begin{tabular}{ccccccc}
  \hline\noalign{\smallskip}
 EFR   & Flux & \multicolumn{5}{c}{Time delay (hour)}  \\
\cline{3-7}
 number & (10$^{19}$ Mx) & 1600 & 304 & 171 & 193 & 211\\
 \hline\noalign{\smallskip}
 20100903F &0.83&1.3&1.7&2.2&2.2&2.1\\
 20111021E &0.90&1.0&1.2& -&1.8&1.5\\
 20110827B &0.91&1.1&1.9&1.1&- &- \\
 20100906C &0.92&1.3&1.7& -&- &-\\
 20100903D &0.98&0.9&1.1&1.3&1.3&1.4\\
 \hline\noalign{\smallskip}

 20100905A &1.1&- &3.8&5.1&5.8&4.8\\

 20110910E &1.28&1.2&1.6&1.9&1.8&1.8\\

 20100902H &1.38&1.8&1.8&1.9&2.5&2.8\\
 20110910D &1.39&0.4&0.7&0.9&1.2&1.0\\
 20110828B &1.48&2.0&2.4&3.1&2.8&3.0\\
 20100903C &1.55&0.5&1.2&3.4&2.2&2.3\\

 20100906D &1.63&1.9&2.6& &3.4&3.3\\

 20100903A &1.75&1.3&2.1&2.3&2.7&2.7\\
 20110828A &1.77&2.5&3.3&4.1&4.1&4.2\\

 20100905D &1.98&1.9&1.7&2.2&3.0&3.2\\

 20111021B &2.1&- & -& -&2.8&2.6 \\
 20100906A &2.61&0.7&0.9&0.7&1.3&1.2\\

 20100902A &2.65&1.7&2.0&2.0&2.1&2.1\\
 20100905E &2.71&2.8&3.9&5.7&5.7&5.7\\
 20100906B &2.83&0.4&0.6&2.0&2.1&2.1\\
 20111021C &3.24&3.1&3.8&- &4.1&4.6\\
 20111021D &3.29&1.8&2.0&1.9&2.5&2.5\\

 20110828C &3.9&2.8&- &- &4.2&3.5\\
 20110910B &5.6&1.1&- &- &3.6&2.7\\
 \hline\noalign{\smallskip}

 20110910A &11.2&5.0&5.9&8.2&6.9&6.6\\

 \noalign{\smallskip}\hline
\end{tabular}
\end{center}
\end{table}

\subsection{Statistical results}
There are 50 EFRs selected to conduct our study in this paper. We downloaded all the corresponding AIA data for the 50 events and found only half of them can give us relatively accurate start time of response, while the others show fluctuant light curves and cannot be used to determine the start time.

Table \ref{Tab:EFRs} gives a list of the selected EFRs and their basic information. All events are listed according to the order of magnetic flux increment. The polarity in table \ref{Tab:EFRs} means the polarity of the EFR, which we used to compute the parameters. Other parameters listed in the table have all been mentioned above. There are two events (20110827D and 20111021D), which we cannot get the separation and separation speed for some reasons.

In the 50 events, seventeen events have fluxes less than 10$^{19}$ Mx, thirty one events have fluxes in the range of (10$^{19}$--10$^{20}$) Mx and only 2 events have fluxes more than 10$^{20}$ Mx. According to the results, the total emerged fluxes are in the range of (0.44--11.2)$\times$10$^{19}$ Mx while the derived flux growth rates are (0.1--1.3)$\times$10$^{19}$ Mx hr$^{-1}$. The maximum area ranges from 17 to 182 arcsec$^2$ and the average magnetic field strength is 29.7 - 116.1 G. The durations of the emergence are between 1 and 12 hours. The maximum distances between the opposite polarities are 7 - 25 arcsec while the separation speeds are from 0.05 to 1.08 km s$^{-1}$.

To check the relationship among the parameters, which describe the property of EFRs, we plot the scatter maps in Figure \ref{fig.mag} and the left panel of Figure \ref{fig.resp}. The result shows that the emerged flux, the area, the average magnetic field strength and the separation are all positively correlated to the emerging duration while the separation speed is negatively correlated to the emerging duration. The emerging rate is positively correlated to the emerged flux.

After preparing the downloaded AIA images, we calculate brightness of each EFR in five wavebands,
\ie, 1600 \AA, 304 \AA, 171 \AA, 193 \AA\ and 211 \AA, and subsequently derive the corresponding
light-curves in order to study the response of the upper solar atmosphere to the relevant flux
emergence. We define the UV/EUV brightening start when the brightness is increased by 20\% of the
maximum enhancement and then compute the time delay of UV/EUV brightening with respect to the
start of flux emergence.

Table \ref{Tab:response} gives a list of EFRs with the corresponding response in the upper atmosphere, of which, twenty percent has fluxes less than 10$^{19}$ Mx while the left has fluxes more than 10$^{19}$ Mx. The response time of the 5 channels are all calculated according to the start of flux emergence. In this way, we get response times in the range of (0.4--5.0), (0.6--5.9), (0.7--8.2), (1.2--6.9), (1.2--6.6) hours in 1600 \AA, 304 \AA, 171 \AA, 193 \AA, and 211 \AA, respectively.

The results are also displayed in Figure \ref{fig.resp} (right panel). It is
shown that the upper atmosphere responds to the flux emergence firstly
in the 1600 \AA\ chromospheric line in half an hour to about 5 hours,
and then tens and hundreds of seconds later, in coronal lines,
such as the 171 \AA\ (T=10$^{5.8}$ K) and 211 \AA\ (T=10$^{6.3}$ K) lines.

\section{Discussion And Conclusions}
\label{sect:sum}

Using the LOS magnetic field data from SDO/HMI, we statistically studied the properties of EFRs
through the seven parameters mentioned above. The inferred relationship of these parameters are generally
consistent with previous results (e.g. Otsuji \etal\ 2007, 2011). We found that
the durations have a larger range (1--12 hours) in our case. All the parameters show a
weak positive correlation with total emerged flux except the separation speed,
which decreases as the emerged flux increases.
This is consistent with the conclusion that tubes of larger EFR are anchored in deeper layers (Javaraiah \etal\ 1997).

Meanwhile, for EFRs with flux less
than 10$^{19}$ Mx in our cases the upper atmosphere does not shown apparent brightness enhancement in coronal lines,
indicating small EFRs tend to interact with low-layer magnetic structure around. However, the heating effect for the lower atmosphere by EFRs with flux less than 10$^{19}$ Mx is not persistent but fluctuant. We deduce that in the lower atmosphere, the magnetic structures around tend to be smaller and lower, when small EFRs which are comparable with surrounding magnetic structures appear, the interaction between them will greatly change the morphologies of both. This will not guarantee the continuously heating.

But when the EFR is larger (here we say with flux greater than 10$^{19}$ Mx), the interaction between the EFR itself and the surrounding magnetic structures may not change the morphology of EFR catastrophically and when it emerges into corona where the surrounding magnetic structures are quite larger and higher, the surrounding magnetic structures may ensure the successive heating. All of these guesses need simulation to prove it out.

In previous papers of Otsuji \etal\ (2007, 2011), they used data from Hinode/SOT which only observes partial area of the sun during a certain time frame. So the parameters that they considered may be not as much as ours, since the data obtained by SOT probably not cover the whole duration of the event , just like missing the beginning phase or didn't contain the phase when magnetic flux reached the maximum. However, we can get the entire information of an EFR as long as we download abundant data thanks to the continuously observing of SDO. Therefor, our statistical result may be more reliable since we have data covering the entire process of the emergence.

The statistical research for response of upper atmosphere has already been studied by Li \etal\ (2007) and Li \& Li (2010), and their works either did not have magnetic information (Li \& Li 2010) or the time resolution was too low (Li \etal\ 2007). In this paper, we have both the LOS magnetogram and five channels of UV/EUV observations for the chromosphere and corona with a time resolution of 12 second. The result shown in Figure \ref{fig.resp} (right panel) manifests that the emerging flux should first reach and heat the chromosphere, and then move to the corona and cause the coronal brightening. It is also mentioned in Li \etal\ (2007) that one could expect that the chromosphere displays enhanced brightening in the Ca II H line earlier than the corona in soft X-ray (SXR), but later than the increase of the integrated magnetic flux. It also manifests that the response time delay is much longer for larger emerged flux. A larger EFR interacts with the surroundings for a longer time and subsequently the heating process lasts a longer time.

It should be mentioned that in our study, we use 12 min cadence HMI LOS magnetic field data,
which is relatively small compared
with the durations of the EFRs and certainly has some effect on our results, which may induce an uncertainty of 12 min for the time delay between flux emergence and upper response. But it is still relatively small when compared with the response time in this paper. The threshold
(20\% of the maximum enhancement) used to define the start time of EUV response could
also slightly affect the time delay. However, these do not change our results as a whole.

In summary, from this statistical study we found that the derived parameters for EFRs
generally have a large range, and all the durations, areas, separations, flux growth rates
and average field strength are in weak positive correlation with the total emerged flux.
EUV emissions are also related to the total emerged flux and delay with respect to
the flux emergence by minutes to hours. The chromosphere responds to flux emergence first and then
the corona. The delayed time increases with the temperature
of the EUV emission, suggesting the successive heating of the upper atmosphere.

\begin{acknowledgements}
This work was supported by the National Natural Science Foundation of China (NSFC, grant number 10873038 and 10833007), the National Basic Research Program of China (2011CB811402). The authors are grateful to the NASA/SDO, the AIA and HMI science teams for the data.

\end{acknowledgements}



\label{lastpage}

\end{document}